# Terahertz Emission from Compensated Magnetic Heterostructures

*Mengji Chen, Rahul Mishra, Yang Wu, Kyusup Lee* and *Hyunsoo Yang\**

Department of Electrical and Computer Engineering and NUSNNI, National University of Singapore, 117576 Singapore
E-mail: eleyang@nus.edu.sg

**Terahertz emission spectroscopy (TES) has recently played an important role in unveiling the spin dynamics at a terahertz (THz) frequency range. So far, ferromagnetic (FM)/nonmagnetic (NM) heterostructures have been intensively studied as THz sources. Compensated magnets such as a ferrimagnet (FIM) and antiferromagnet (AFM) are other types of magnetic materials with interesting spin dynamics. In this work, we study TES from compensated magnetic heterostructures including CoGd FIM alloy or IrMn AFM layers. Systematic measurements on composition and temperature dependences of THz emission from CoGd/Pt bilayer structures are conducted. It is found that the emitted THz field is determined by the net spin polarization of the laser induced spin current rather than the net magnetization. The temperature robustness of the FIM based THz emitter is also demonstrated. On the other hand, an AFM plays a different role in THz emission. The IrMn/Pt bilayer shows negligible THz signals, whereas Co/IrMn induces sizable THz outputs, indicating that IrMn is not a good spin current generator, but a good detector. Our results not only suggest that a compensated magnet can be utilized for robust THz emission, but also provide a new approach to study the magnetization dynamics especially near the magnetization compensation point.**

# 1. Introduction

During the past decades, THz technologies have been studied intensively,[1-4] for a wide range of promising applications such as the chemical composition analysis,[5, 6] integrated circuits failure analysis,[7] and spintronics characterization.[8] In order to ensure a high performance of the THz module in these applications, it is highly desired to develop efficient THz components such as the THz emitter,[9-13] detector,[14] phase shifter,[15] and modulator.[16] In particular, the development of an efficient THz emitter is considered as a key challenge in the field of THz technology.[1] At present, in a femtosecond laser driven system, photoconductive antenna,[10] electro-optic crystal,[12, 13] and plasma[11, 17] have been widely applied as THz emitters for different kinds of THz research and applications. However, there are various shortcomings in the exsiting THz emitters such as a low signal-noise ratio, high cost and narrow bandwidth. Therefore, an efficient and robust THz source with a low cost is still in high demand.

Spintronics materials have been utilized to construct efficient and low cost THz emitters. In 2004, the first THz emission from a laser induced ultrafast demagnetization of Ni was reported.[18] It was after a decade in 2013 that spin-orbit interaction based terahertz emission was demonstrated from the FM/NM heterostuctures.[19] A fs laser was used to excite an ultrafast spin current $j_s$ in the FM layer. The spin current diffuses into the NM layer, where it is converted into an charge current $j_c$ due to the inverse spin-Hall effect (ISHE). Finally, this ultrafast charge current leads to a radiation of the electromagnetic THz signal.[19] This new approach has ignited interest in development of efficient THz emitters based on FM/NM heterostructures.[20-23]

Antiferromagnetically coupled magnetic materials such as a FIM and AFM have the ultrafast dynamics in the range of THz, minimal stray fields due to the reduced magnetization, a resistance against external magnetic field perturbation, and a high thermal stability due to their ultrahigh anisotropy.[24, 25] In this work, we study THz emission from different heterostructures comprising of FIM and AFM materials. A strong THz output is observed from a compensated FIM/NM bilayer, $Co_{74}Gd_{26}$ (7 nm)/Pt (6 nm) with nearly zero magnetization, while there is no measureable THz signal from an AFM/NM bilayer, $Ir_{25}Mn_{75}$ (7 nm)/Pt (6 nm). However, when the AFM is used a spin detector in an AFM/FM bilayer, $Ir_{25}Mn_{75}$ (6 nm)/Co (3 nm), a sizable THz signal is obtained. In addition, composition and temperature dependent studies of THz emissiom are performed on the FIM/NM bilayer. It is found that the net spin polarization, rather than the net magnetization, plays a dominant role in the THz emission. A FIM/NM bilayer is able to sustain a temperature cycle of 473 K. Our works not only provide a better understanding of compensated magnet based THz emission but also open up a possibility of ultilizing them as a practical THz emitter.

## 2. Sample Preparation and Measurement Results

For this work, thin films are deposited on the glass substrates by magnetron sputtering with a base pressure less than $5 \times 10^{-9}$ Torr. The FIM layer is deposited by co-sputtering of Co and Gd and the AFM layer is deposited by sputtering of $Ir_{25}Mn_{75}$ target. All the films are protected by a 3 nm thick $SiO_2$ layer to prevent oxidization. The magnetic property of the samples was measured by a vibrating sample magnetometer (VSM). **Figure 1**a shows the schematic of the TES system, which is based on a stroboscopic setup. We employ a laser source with a full width at the half maximum of 120 fs, a center wavelength of 800 nm and a repetition rate of 1 kHz. A linearly polarized pump laser pulse with an energy of 220 µJ is normally incident from the sample layer side, with an external magnetic field ~1000 Oe applied along the *-y* direction in Figure 1b. With a 1 mm thick ZnTe (110) crystal acting as a THz detector, the ellipticity of the probe laser pulse with an energy of 2 µJ is modulated by the THz electric field due to the electro-optical effect. Based on the electro-optic sampling method, the THz signal is recorded in the time domain.

We first study the THz emission from the FIM/NM bilayer. A typical sample structure used for the experiments is shown in Figure 1b. A 7 nm FIM $Co_{1-x}Gd_x$ layer is deposited on the glass substrate with a 6 nm NM Pt layer on top. Various $Co_{1-x}Gd_x$ samples with different compositions varying from x = 0 to 55 are prepared. **Figure 2**a shows the saturation magnetization of the samples measured by a VSM. The compensation point is found to be x = ~26. The samples for x < 26 are Co rich in which the magnetic moments of Co sublattice align parallel to the external magnetic field, and those of Gd align antiparallel to the external magnetic field. The samples with x > 26 are Gd rich with the magnetic moments of Co sublattice aligning antiparallel to the external magnetic field and those of Gd are parallel to the field. TES measurement results are shown in Figure 2b. The intensity of THz signal keeps decreasing as the Gd fraction increases, while the saturation magnetization $M_s$ firstly decreases then increases after compensation point (x = 26). The $M_s$ reaches almost zero at x = 26, however the THz signal has an abrupt sign change around x = 26 without showing a fully quenched THz signal. Therefore, it seems that the net magnetization is not strongly correlated to the emitted THz signal.

In order to better understand the THz emission from nearly compensated structures, we compare THz emission from a nearly compensated FIM/NM bilayer (x = 26), $Co_{74}Gd_{26}$ (7 nm)/Pt (6 nm), with the AFM/NM bilayer structure, $Ir_{25}Mn_{75}$ (7 nm)/Pt (6 nm), as shown in **Figure 3,** in which both samples have zero net magnetization. Based on the theoretical model of spin charge conversion in the THz emission,[19] $\boldsymbol{j_c} = \gamma \boldsymbol{j_s} \times \mathbf{M}/|\mathbf{M}|$, where $\gamma$ is the effective spin Hall angle and **M** is the sample magnetization, it is expected that both the nearly compensated FIM/NM and AFM/NM bilayers should produce negligible THz signals. As expected, no measurable THz signal is detected from the IrMn/Pt bilayer in Figure 3. However, a nearly compensated FIM/NM bilayer structure, $Co_{74}Gd_{26}$/Pt acts as an efficient THz source. It was observed that the sign of THz signal from the $Co_{74}Gd_{26}$/Pt bilayer reverses upon reversing either the external magnetic field direction (+H or –H) or the sample around the *x*-axis (not shown). The above results pinpoint that the observed THz signal from $Co_{74}Gd_{26}$/Pt has a spin-current-based origin from the heterostructure itself, rather than the demagnetization dynamics contribution.[18, 26] The very different results from the nearly compensated FIM/NM and AFM/NM suggest that the generated spin current is negligible in AFM/NM, whereas the spin current from a FIM/NM is finite even though the net magnetization is close to zero.

The THz emission from the nearly compensated FIM/NM bilayer can be understood from the localized and delocalized nature of the electrons carrying magnetic moments in Gd and Co, respectively. The spin-split bands of Gd sublattice originate from the 4*f* electrons which are spatially localized below the Fermi level about 8 eV and are difficult to be excited by the laser pulse. On the other hand, the spin-split bands of Co sublattice are much closer to the Fermi level (~1 eV) and are easily excited when the laser pulse thermalizes the CoGd layer. Thus, the super diffusive spin current generated in the FIM layer is dominated by the electrons from the Co sublattice, and accounts for the THz emission from the nearly compensated CoGd/Pt bilayer.[27-31] It can also be concluded that the net spin polarization rather than net magnetization is mainly responsible for the THz emission from a magnetic/heavy metal heterostructure. The decreasing THz emission with increasing the Gd composition as well as the sign change of THz signal in Figure 2b matches well with the spin polarization values of CoGd.[31]

While there is no THz generation due to the lack of a spin current in the IrMn/Pt bilayer with the laser excitation, an AFM is reported as an effective spin detector.[32, 33] A typical THz signal from an AFM/FM bilayer, $Ir_{25}Mn_{75}$ (6 nm)/Co (3 nm) is shown in **Figure 4**a. With a spin current in Co excited by a *fs* laser pump, the IrMn layer converts the spin current to the charge current, leading to the THz emission. On the other hand, Figure 4a also shows that there is no detectable THz signal generated from an AFM/NM bilayer, $Ir_{25}Mn_{75}$ (6 nm)/Pt (3 nm) because of a lack of the spin current source. We further perform the thickness dependent study with a sample structure of glass substrate/$Ir_{25}Mn_{75}$ (0 − 10 nm)/Co (3 nm)/$SiO_2$ (3 nm), as shown in Figure 4b. With 1 nm IrMn with the Co layer, the THz signal shows a peak intensity, and then the THz signal becomes gradually weaker as we increase the thickness of IrMn. The above thickness dependent results can be attributed to the combined effect of the spin diffusion in IrMn, the THz absorption in the IrMn/Co structure and the Fabry-Perot effect.[20, 22] When the thickness of the IrMn layer is below the spin diffusion length, the ultrafast spin current from the Co layer cannot be fully converted into the charge current, resulting in a smaller THz signal. In the previous study,[20] the Fabry-Perot interference effect is considered in the thickness dependence of THz emission, which is believed to enhance both the laser pump and the emitted THz waves resonantly. However, in our thickness dependent measurement, the THz signal decreases with increasing the IrMn thickness and no critical thickness, in which the maximum THz signal appears[20, 22], is observed. Therefore, we believe a significant THz absorption in the IrMn layer suppresses the enhancement from the Fabry-Perot effect and the THz absorption in the IrMn layer contributes more significantly in our thickness dependent study. As a result, the peak THz signal should be observed with a thickness of IrMn near the spin relaxation length, due to the maximum spin charge conversion and less THz absorption from the thin film. Our observation is consistent to the reported spin relaxation length of $Ir_{25}Mn_{75}$ (< 1 nm).[33, 34]

In a FIM, the magnetization compensation temperature ($T_{comp}$) also plays an important role in the magnetization configuration and thus spin polarization. A Gd rich sample ($Co_{63}Gd_{37}$/Pt) is chosen for a temperature dependent study of THz emission, as shown in **Figure 5**a. A sign reversal of the THz signal near 433 K is observed and the intensity of THz emission becomes smaller when the temperature is closer to 433 K. Compared with the temperature dependence data from VSM shown in Figure 5b, a sign reversal of the THz signal is observed across $T_{comp}$. The magnetic alignment of the Co sublattice as well as the spin polarization change its direction across $T_{comp}$,[31] leading to the inverted THz emission sign in line with the previous works.[35, 36] Different from the composition dependent results, the THz signal decrease substantially around the compensation temperature. This phenomenon can be attributed to thin film inhomogeneity in the deposition process or sputter target, which induces the nonuniform concentration profiles of thin film with the mix of Co and Gd rich regions.[31, 37, 38] Near the compensation temperature, although the sample has minimum magnetization, some regions of the samples are Co rich and some are Gd rich due to inhomogeneity. As the net spin polarization is decided by Co, the sample near compensation ends up having a small polarization as the Co polarization in different regions cancels each other. A small temperature difference, where the THz signal reverses and VSM data show the minimum value, can be attributed to the laser induced heating of the sample.[27, 35]

The temperature robutness is critical for real THz applications, and a finite THz signal above 400 K in Figure 5a indicates a potential for applications. Thus, we investigate the performance of the FIM/NM based emitters after heating to a high temperature. For the case of $Co_{63}Gd_{37}$/Pt bilayer structure (Gd rich sample) shown in **Figure 6**, the THz signal reverses at ~473 K, which is above $T_{comp}$. Subsequently, when the sample is cooled down back to the room temperature, the THz signal can be fully recovered. Therefore, a FIM/NM system demonstrates a non-hysteretic temperature cycle with respect to the THz signal after heating up and cooling down to the same temperature.

## 3. Conclusion

We study THz emisison from compensated magnet based bilayer structures with a *fs* laser excitation. We show that the net spin polarization, rather than the net magnetization, accounts for the THz generation. With a FIM acting as a spin source, a sizable THz output is observed dominated by Co sublattices. It is found that AFM IrMn plays a role as an effective spin sink rather than a spin source in THz emission. In addition, an non-hysteretic temperature cycle property makes this the FIM/NM THz emitter an alternative THz source.

**Acknowledgements**
This research is supported by the National Research Foundation (NRF), Prime Minister's Office, Singapore, under its Competitive Research Programme (CRP Award No. NRF CRP12-2013-01).

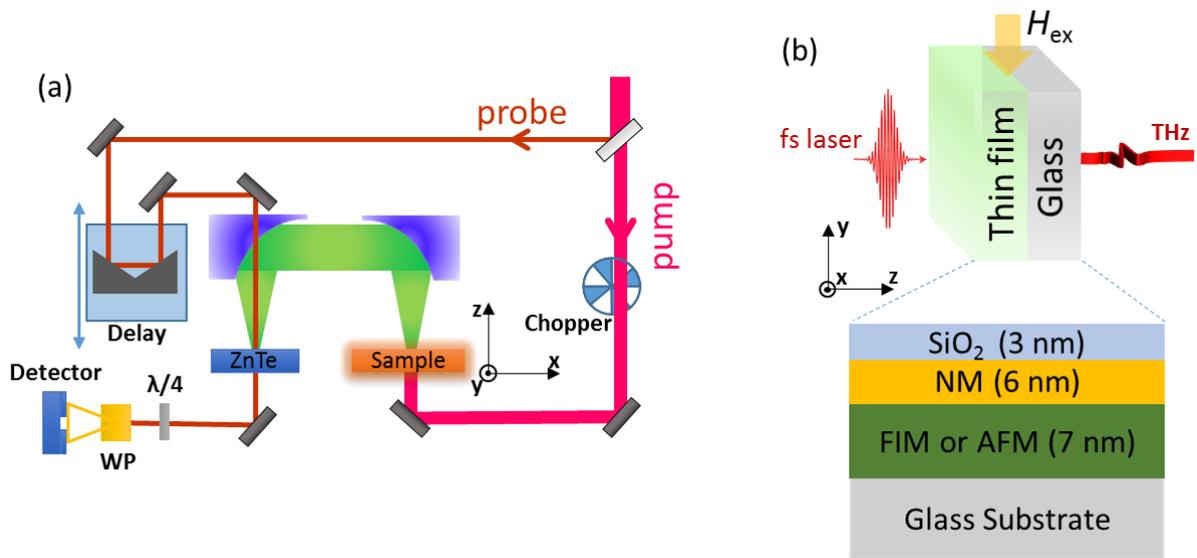

**Figure 1.** a) Schematics of the THz emission spectroscopy setup. b) Schematics of film stack and THz emission geometry. An 800 nm polarized laser pump is normally incident from the thin film side and emitted THz signal is measured along the *x* axis. An external magnetic field ($H_{ex}$ = 1000 Oe) is applied along the −*y* axis.

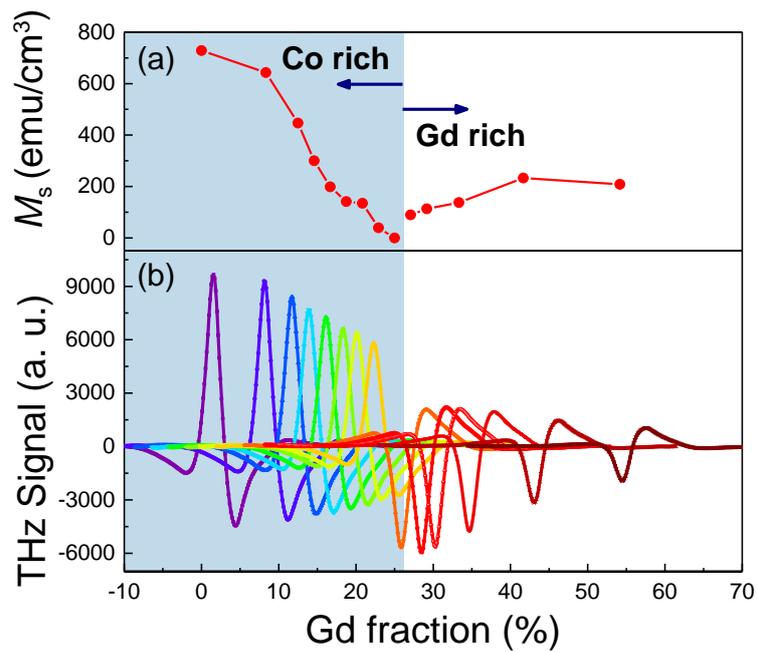

**Figure 2.** a) The saturation magnetization measured by a VSM. b) Composition dependence of the THz signal of CoGd/Pt samples as a function of Gd fraction. The shaded area indicates the Co rich region and the white background is for Gd rich samples. All the THz signals have the same time delay, but are shifted horizontally according to the corresponding Gd fraction. The maximum peak or dip location of each THz curve indicates the Gd fraction.

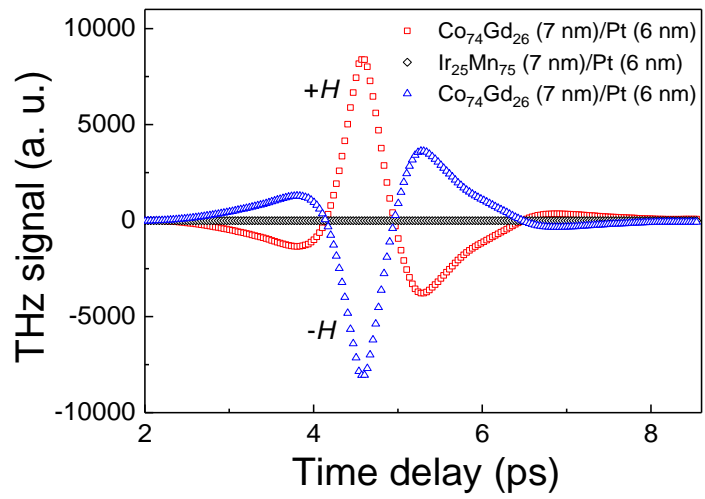

**Figure 3.** Typical THz signal from the nearly compensated FIM/NM and AFM/NM bilayers. For the FIM/NM bilayer opposite external magnetic fields (+$H$ and −$H$) are applied.

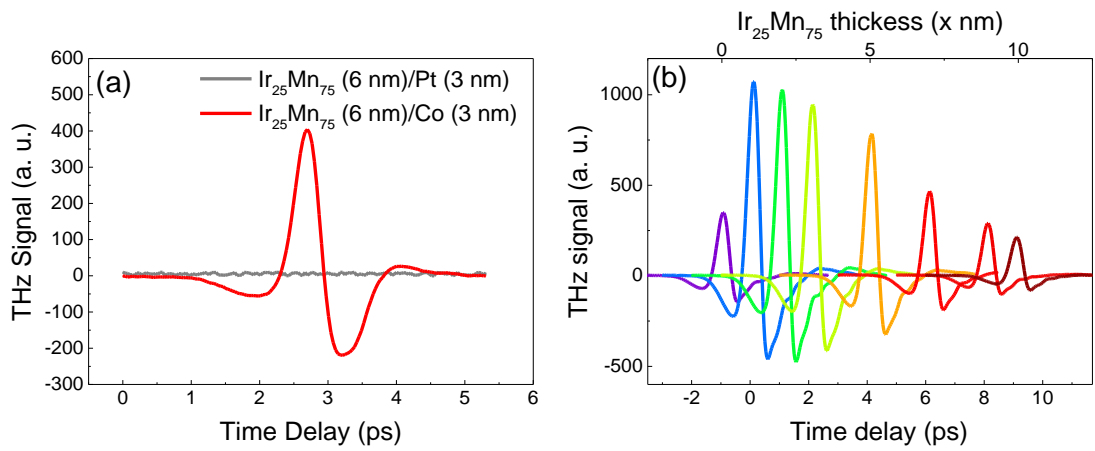

**Figure 4.** a) Typical THz signal from AFM/NM and AFM/FM bilayers. An external magnetic field (~1000 Oe) is applied. b) IrMn thickness dependence of THz emission from glass substrate/Ir$_{25}$Mn$_{75}$ (0 - 10 nm)/Co (3 nm)/SiO$_2$ (3 nm). All the THz signals have the same time delay, but are shifted horizontally according to the IrMn thickness. The peak location of each THz curve indicates the IrMn thickness.

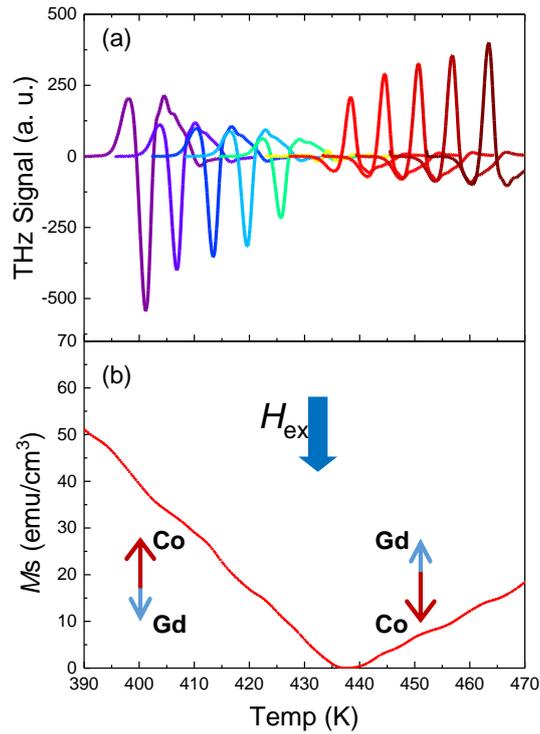

**Figure 5.** a) Temperature dependence of the emitted THz signal from the $Co_{63}Gd_{37}$/Pt bilayer. All the THz pulses have the same time delay, but are shifted horizontally with respect to the corresponding temperature. The maximum peak or dip location of each THz curve indicates the measured temperature. b) The saturation magnetization measured with a VSM as a function of temperature. The magnetization direction of Co and Gd is indicated with respect to the external magnetic field direction ($H_{ex}$).

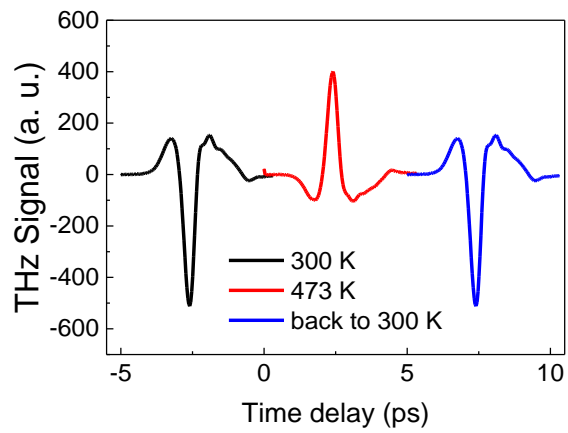

**Figure 6.** a) Emitted THz signal from the $Co_{63}Gd_{37}$/Pt bilayer at different temperatures. The black line shows the data at room temperature (300 K), followed by a measurement at 473 K shown by the red line. The blue line shows the measurement data at room temperature after cooling down to room temperature from 473 K. THz pulses are shifted horizontally for clarity.

**An efficient THz emission from compensated ferrimagnet/nonmagnetic metallic heterostructures is demonstrated.** The THz emission is determined by the net spin polarization rather than the net magnetization. Antiferromagnet plays a role as an effective spin sink rather than a spin source in THz emission. Ferrimagnet based THz emitters are promising not only for THz applications, but also offering an opportunity to study the magnetization dynamics.

**Keyword**
THz emission, ferrimagnet, antiferromagnet, spintronics, magnetic heterostructure

*Mengji Chen, Rahul Mishra, Yang Wu, Kyusup Lee* and *Hyunsoo Yang\**

**Terahertz Emission from Compensated Magnetic Heterostructures**

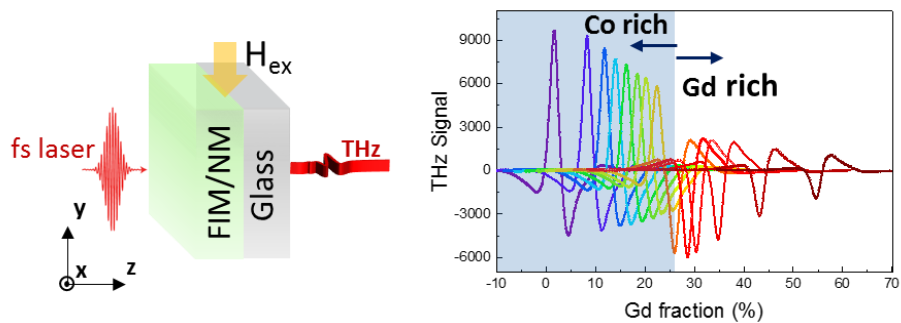